\documentstyle[12pt]{article}

\catcode`\@=11
\long\def\@makefntext#1{
\protect\noindent \hbox to 3.2pt {\hskip-.9pt
$^{{\ninerm\@thefnmark}}$\hfil}#1\hfill}		

\def\@makefnmark{\hbox to 0pt{$^{\@thefnmark}$\hss}}  

\def\ps@myheadings{\let\@mkboth\@gobbletwo
\def\@oddhead{\hbox{}
\rightmark\hfil\ninerm\thepage}
\def\@oddfoot{}\def\@evenhead{\ninerm\thepage\hfil
\leftmark\hbox{}}\def\@evenfoot{}
\def\sectionmark##1{}\def\subsectionmark##1{}}

\renewcommand{\thefootnote}{\fnsymbol{footnote}}

\newcounter{sectionc}\newcounter{subsectionc}\newcounter{subsubsectionc}
\renewcommand{\section}[1] {\vspace*{0.6cm}\addtocounter{sectionc}{1}
\setcounter{subsectionc}{0}\setcounter{subsubsectionc}{0}\noindent
	{\normalsize\bf\thesectionc. #1}\par\vspace*{0.4cm}}
\renewcommand{\subsection}[1] {\vspace*{0.6cm}\addtocounter{subsectionc}{1}
	\setcounter{subsubsectionc}{0}\noindent
	{\normalsize\it\thesectionc.\thesubsectionc. #1}\par\vspace*{0.4cm}}
\renewcommand{\subsubsection}[1]
{\vspace*{0.6cm}\addtocounter{subsubsectionc}{1}
	\noindent
{\normalsize\rm\thesectionc.\thesubsectionc.\thesubsubsectionc.
	#1}\par\vspace*{0.4cm}}

\newcounter{appendixc}
\newcounter{subappendixc}[appendixc]
\newcounter{subsubappendixc}[subappendixc]

\renewcommand{\appendix}[1] {\vspace*{0.6cm}
        \refstepcounter{appendixc}
        \setcounter{figure}{0}
        \setcounter{table}{0}
        \setcounter{equation}{0}
        \renewcommand{\thefigure}{\Alph{appendixc}.\arabic{figure}}
        \renewcommand{\thetable}{\Alph{appendixc}.\arabic{table}}
        \renewcommand{\theappendixc}{\Alph{appendixc}}
        \renewcommand{\theequation}{\Alph{appendixc}.\arabic{equation}}
        \noindent{\bf Appendix \theappendixc #1}\par\vspace*{0.4cm}}

\def\abstracts#1{{

\centering{\begin{minipage}{12.2truecm}\footnotesize\baselineskip=12pt\noindent
	\centerline{\footnotesize ABSTRACT}\vspace*{0.3cm}
	\parindent=0pt #1
	\end{minipage}}\par}}

\newcounter{itemlistc}
\newcounter{romanlistc}
\newcounter{alphlistc}
\newcounter{arabiclistc}

\newcommand{\fcaption}[1]{
        \refstepcounter{figure}
        \setbox\@tempboxa = \hbox{\footnotesize Fig.~\thefigure. #1}
        \ifdim \wd\@tempboxa > 6in
           {\begin{center}
        \parbox{6in}{\footnotesize\baselineskip=12pt Fig.~\thefigure. #1}
            \end{center}}
        \else
             {\begin{center}
             {\footnotesize Fig.~\thefigure. #1}
              \end{center}}
        \fi}

\newcommand{\tcaption}[1]{
        \refstepcounter{table}
        \setbox\@tempboxa = \hbox{\footnotesize Table~\thetable. #1}
        \ifdim \wd\@tempboxa > 6in
           {\begin{center}
        \parbox{6in}{\footnotesize\baselineskip=12pt Table~\thetable. #1}
            \end{center}}
        \else
             {\begin{center}
             {\footnotesize Table~\thetable. #1}
              \end{center}}
        \fi}

\def\@citex[#1]#2{\if@filesw\immediate\write\@auxout
	{\string\citation{#2}}\fi
\def\@citea{}\@cite{\@for\@citeb:=#2\do
	{\@citea\def\@citea{,}\@ifundefined
	{b@\@citeb}{{\bf ?}\@warning
	{Citation `\@citeb' on page \thepage \space undefined}}
	{\csname b@\@citeb\endcsname}}}{#1}}

\newif\if@cghi
\def\cite{\@cghitrue\@ifnextchar [{\@tempswatrue
	\@citex}{\@tempswafalse\@citex[]}}
\def\citelow{\@cghifalse\@ifnextchar [{\@tempswatrue
	\@citex}{\@tempswafalse\@citex[]}}
\def\@cite#1#2{{$\null^{#1}$\if@tempswa\typeout
	{IJCGA warning: optional citation argument
	ignored: `#2'} \fi}}

 1
 1
 1

\font\ninerm=cmr9

\begin{document}

\newcommand{\st}{\scriptstyle}
\newcommand{\sst}{\scriptscriptstyle}
\newcommand{\mco}{\multicolumn}
\newcommand{\epp}{\epsilon^{\prime}}
\newcommand{\vep}{\varepsilon}
\newcommand{\ra}{\rightarrow}
\newcommand{\ppg}{\pi^+\pi^-\gamma}
\newcommand{\vp}{{\bf p}}
\newcommand{\ko}{K^0}
\newcommand{\kb}{\bar{K^0}}
\newcommand{\al}{\alpha}
\newcommand{\ab}{\bar{\alpha}}
\def\be{\begin{equation}}
\def\ee{\end{equation}}
\def\bea{\begin{eqnarray}}
\def\eea{\end{eqnarray}}
\def\CPbar{\hbox{{\rm CP}\hskip-1.80em{/}}}

\centerline{\normalsize\bf BOUNDS ON $\Delta B=1$ COUPLINGS}
\centerline{\normalsize\bf  IN THE
SUPERSYMMETRIC STANDARD MODEL}
\baselineskip=16pt

\vspace*{0.6cm}
\centerline{\footnotesize MARC SHER}
\baselineskip=13pt
\centerline{\footnotesize\it Department of Physics, College of William and
Mary}
\baselineskip=12pt
\centerline{\footnotesize\it Williamsburg VA  23185, USA}
\centerline{\footnotesize E-mail: sher@wmheg.physics.wm.edu}
\vspace*{0.3cm}
\centerline{\footnotesize and}
\vspace*{0.3cm}
\centerline{\footnotesize J.L. GOITY}
\baselineskip=13pt
\centerline{\footnotesize\it Department of Physics, Hampton University, Hampton
VA 23668, USA }
\centerline{\footnotesize\it and}
\centerline{\footnotesize\it CEBAF, 12000 Jefferson Ave., Newport News, VA
23606, USA}
\centerline{\footnotesize E-mail:  goity@cebaf.gov}
\def\lm{\lambda}
\def\ov{\overline}
\vspace*{0.9cm}
\abstracts{
The most general supersymmetric model contains baryon number violating terms of
the form $\lm_{ijk}\ \ov{D}_i\ \ov{D}_j\ \ov{U}_k$ in the superpotential.  We
reconsider the bounds on these couplings, assuming that lepton number
conservation ensures proton stability.  These operators can mediate $n-\ov{n}$
oscillations and double nucleon decay.  We show that neutron oscillations do
not, as previously claimed, constrain the $\lm_{dsu}$ coupling; they do provide
a bound on the $\lm_{dbu}$ coupling, which we calculate.  We find that the best
bound on $\lm_{dsu}$ arises from double nucleon decay into two kaons.  There
are
no published limits on this process; experimenters are urged to examine this
nuclear decay mode.  Finally, the other couplings can be bounded by the
requirement of perturbative unification. }

\vspace*{0.6cm}
\normalsize\baselineskip=15pt
\setcounter{footnote}{0}
\renewcommand{\thefootnote}{\alph{footnote}}

In the standard electroweak model, conservation of baryon number and lepton
number arises automatically from gauge invariance.  This is not the case in
supersymmetric models, however.  In the most general low-energy supersymmetric
model, one has terms which violate lepton number and terms which violate baryon
number.  Since the presence of both of these may lead to unacceptably rapid
proton decay, one or both must generally be suppressed by a discrete symmetry.
In the most popular model, R-parity, given by $(-1)^{3B+L+F}$ is imposed,
leading to baryon and lepton number conservation.  However, there is no {\it a
priori} reason that R-parity must be imposed, it is quite possible that only
one
of the quantum numbers is conserved.  There has been extensive discussion of
the
possibility that lepton number is violated, but relatively little investigation
of the possibility that lepton number is conserved and baryon number is
violated.

In this case, a term will appear in the superpotential given by
$\lm_{ijk}\ \ov{D}_i\ \ov{D}_j\ \ov{U}_k$ where the indices give the generation
number and the chiral superfields are all righthanded isosinglet antiquarks.
Since the term is symmetric under exchange of the first two indices, and is
antisymmetric in color, it must be antisymmetric in the first two flavor
indices, leaving nine couplings which will be designated (in an obvious
notation) as $\lm_{dsu}, \lm_{dbu}, \lm_{sbu}, \lm_{dsc}, \lm_{dbc},
\lm_{sbc},
\lm_{dst}, \lm_{dbt}$ and $\lm_{sbt}$.

There are many models in which low energy baryon number is violated and yet
lepton number is conserved; the most familiar are some left-right symmetric
models.  It is widely believed that the strongest bound on B-violating
operators comes from neutron oscillations, which violate B by two units.  The
first discussion on the effects of some of these operators in supersymmetric
models used neutron oscillations to bound $\lm_{dsu}$ and $\lm_{dbu}$; this
result remains widely cited today.

In this talk, I will discuss the most stringent bounds that can be placed on
these nine couplings.  First, it will be pointed out, in contrast with previous
claims, that neutron oscillations do {\it not} provide any significant bound at
all on the $\lm_{dsu}$ coupling, due to a suppression factor which was
neglected
in the original calculation.  This suppression is less severe for the
$\lm_{dbu}$
coupling, and we will obtain a bound in that case.  It will then be pointed out
that the strongest bound on the $\lm_{dsu}$ coupling will come from limits on
double nucleon decay (in a nucleus) into two kaons of identical strangeness,
and
will estimate the bound.  Finally, the recent work of Brahmachari and Roy noted
that the
$\lm_{dbt}$ and $\lm_{sbt}$ couplings can be bounded by requiring perturbative
unification; we will extend their work to cover all of the additional
couplings.  This work will appear in Physics Letters B in March or April of
1995; the reader is referred there for the details of the calculation and for a
list of references (which are neglected here due to space limitations).

In the first attempt to place a bound on some of the above operators considered
the effect of the $\lm_{dsu}$ and $\lm_{dbu}$ terms on neutron oscillations
through the process $(udd \rightarrow
\tilde{\ov{d}}_id\rightarrow \tilde{g}\rightarrow \tilde{d}_i\ov{d}
\rightarrow\ov{u}\ \ov{d}\ \ov{d})$, where $\tilde{d}$ is a squark and
$\tilde{g}$ is a gluino.  The effects of intergenerational mixing were included
by putting in an arbitrary mixing angle, assigned a value of 0.1.  However,
there is a much more severe suppression factor which results in this process
giving no significant contribution to neutron oscillations.

Consider the term $\lm_{dsu}\ov{D}\ \ov{S}\ \ov{U}$.  It violates B by one unit
and strangeness (S) by one unit.  However, it conserves B-S.  Since neutron
oscillations violate B but not S, strangeness violation must appear elsewhere
in
the diagram.  This means that there must be flavor-changing electroweak
interactions (involving either a W or a charged Higgs or chargino) in the
diagram.  Since only isodoublets participate in the weak interactions, and the
baryon number violating term has only isosinglets, there must be mass
insertions
(on the squark lines)--at least two in the simplest case.  Thus, there will be
electroweak interactions in the diagram, and an additional suppression factor
of
$m^2_s/m^2_W$, where $m_s$ is the strange quark mass.  This makes the
contribution of the $\lm_{dsu}$ term highly suppressed. (Note:  off-diagonal
gluino couplings are generally much smaller, and will not contribute
significantly).

The contribution involving the $\lm_{dbu}$ terms will only be suppressed by
$m^2_b/m^2_W$, which is not negligible.  The leading contribution is from a
box diagram in which a $\tilde{b}_L$ and a $\ov{d}_L$ turn into a
$\tilde{\ov{b}}_L$ and a $d_L$; the $\tilde{b}_L$ arises from the $\lm_{dbu}$
coupling with a mass insertion on the squark line.  We have calculated this
contribution (see the Physics Letters paper for details), and find a bound on
$\lm_{dbu}$ which varies from $.002$ and $.1$ as the squark mass varies from
200
GeV to 600 GeV.  For squark masses above a TeV, no useful bound emerges.

The best bound on the $\lm_{dsu}$ term arises from double nucleon decay into
two
kaons, which violates both B and S, but not B-S.  Thus, mass insertions and
electroweak interactions are unnecessary.  Here, the diagram simply involves
having two $\lm_{dsu}$ interactions turns $u$ and $d$ quarks into $s$
anti-squarks, which exchange a gluino and turn into $s$ antiquarks; the
spectators then form the two kaons. Taking the limit on the rate to be
$10^{30}$ years, we find the bound to be $\lm_{dsu}< 10^{-15}R^{-5/2}$, where
$R$ is roughly the ratio of a hadronic scale to the squark mass; for a hadronic
scale of 300 MeV and a squark mass of 300 GeV, this bound is $10^{-7}$. A
similar bound was obtained by assuming that a neutron could oscillate into a
$\ov{\Xi}$, which then annihilates with another neutron in the nucleus to
produce two kaons.  Note that there is currently no published bound on this
decay; experimenters at Kamiokande are currently analysing the possibility,
by considering the decay of the oxygen nuclei in their detector to carbon-14
and two charged  kaons.

What about the other seven couplings?  It was recently noted by Brahmachari and
Roy that in a unified theory precise bounds can be obtained by requiring that
the couplings be perturbative up to the unification scale.  We have generalized
their results (which applied only to two of the couplings) to all seven, and
find an upper bound of $1.25$ on each of them (the bound actually applies to
the
square root of the sum of the squares of combinations of three of the
couplings,
so is a bit stronger), with a slightly stronger bound for couplings involving
the top quark.  Note that if the fixed point is saturated for one of these
couplings (as may be the case for the top quark Yukawa coupling), then this
bound would, in fact, be the value of the coupling at low energy, in which case
the scalar quarks  would have to be heavier than the top quark in order to
avoid the domination of top quark decays into a light quark and a squark.

We are grateful to Herbi Dreiner for many enlightening discussions and
references, and Rabi Mohapatra for helpful remarks concerning neutron
oscillations.  The talk was presented by Marc Sher

\end{document}